# UVGI Scientific Calculator

Nicolas Bouri, Vladimir Shatalov[*]

*Covid Clean, Minneapolis, Minnesota, U.S.*

The physical basis and algorithm of the ultraviolet germicidal irradiation (UVGI) scientific calculator are presented herein. The algorithm was implemented in the web application *UVGI Scientific Calculator* (found at https://covid-19-clean.org/calculator-journal-report/) that has been used to facilitate the engineering and design of air-purifying equipment being developed for use in hospitals, commercial and residential applications. Presented examples of the calculations illustrate how different factors of the construction of a UVGI air-purifier may influence the degree of virus inactivation.

## Introduction

In this paper, we present and explain the physics underpinning the development of a novel ultraviolet germicidal irradiation (UVGI) scientific calculator. The design of this *UVGI scientific calculator* allows ready and quick access to estimations of ultraviolet germicidal irradiation (UVGI) in a broad range of environments. UVGI band ultraviolet C (UVC) radiation between 200-280nm is a particularly harmful wavelength to microorganisms and cell structures.

Starting from the pioneering work of Downes and Blunt (1878) [1], UVGI technologies have been validated as a means to reduce airborne biohazards in indoor environments. The application of UVGI disinfection has been an accepted practice since the mid-20[th] century and is widely used today in medical sanitation and for other sterilization purposes. Publications concerning the history, methods of operation, and effectiveness of these applications exist (see, for example, [2]). This paper contains a description of the calculation steps of the ultraviolet germicidal irradiation air-purifying temporary negative pressure isolation (UVGI-AP-TNPI) system, sources of deactivation doses, and reflectivity of walls.

The model described in the present research is based on an irradiation chamber with various reflective materials and a UVGI source along the interior central axis. Biological deactivation rates are calculated based on several parameters, which include airflow, distance, time, radiant power, and chamber reflectivity. Biohazardous air enters the ultraviolet germicidal irradiation air-purifier (UVGI-AP) and is released to the next environment. In this case, the patient and staff are isolated from one another in both directions—supply and exhaust.

---

[*] *Corresponding author: e-mail: vladishat@gmail.com*

Most of studies on the UVGI topic have focused on the upper room ultraviolet germicidal irradiation (UR-UVGI) or open room cleaning with static samples held in a petri dish. Through the data on the combination of UVGI research we can conclude UVGI is highly effective at biohazard deactivation in air, water and on surfaces. Medical facilities that use UVGI to disinfect rooms show significant sanitation improvements over traditional cleaning methods [3].

Nicolas Bouri created and holds two provisional patents [4] on a novel UVGI-AP-TNPI that allowed the UVGI disinfection of a hospital room with a mobile and modular TNPI unit in the center by utilizing UV stabilized polycarbonate that blocks all electromagnetic wavelengths under 400nm. This increases the effectiveness of room sanitization without removing the patient. This allows for quicker turnover and the possible use during surgeries. The system utilizes UVGI-AP's on both supply and intake. This provides clean air to the patient and cleanses the air escaping the isolation unit. What was not known was the effectiveness of the air cleaning process.

The calculator presented clear problems with the efficiency of traditional UVGI-AP's and were shown to be highly ineffective. With this insight highly reflective materials in the UV-C range were investigated. The porous polytetrafluoroethylene (PTFE) was found to be the most reflective at 254nm where UVGI is commonly used. Used in a cylindrical chamber without material that absorbs UV-C (for example all metals, aluminum, steel, and other materials such as filters), the efficiency grew ten-fold over steel. For the first time in the industry, this allowed for proof up to 99.99% efficiency with airflows rates required for isolation rooms. The minimum air-changes per hour (ACH) required is 12. Solutions used today still require the ventilation of the escaping air to a heating ventilation and air conditioning (HVAC) system or out a window.

The design of the system allows the air to be vented directly back into the outside room without safety concerns. Given the mobile and modular nature of the systems, simple construction and ability to allow visible light, infrared, cell phone communications, medical wireless telemetry, radio frequency and wireless energy to transmit through the shell significant global capabilities to contain air-borne biohazard are increased. The low cost and mobility of the systems allow for substantial increases in isolation units not only in the US but globally and can mitigate pandemics. Nicolas Bouri started Covid Clean back in March to assist with the Covid-19 pandemic and was incorporated on April 8$^{th}$, 2020.

The mathematical model of inactivation used elsewhere was static and needed to be redone every time the parameters changed to chase increased efficiency. The process was slow and tedious therefor the calculator presented in this paper was created. It was important to calculate the changes in real time as parameters changed. The real time changes can be monitored in the UVGI-AP's and if safety parameters are exceeded the system can adjust airflow or power to ensure safe operation. If not, the system can issue warnings and can be monitored wirelessly and remotely.

## Method

Survival rate $S(\vec{r}, t)$ in any point of the irradiated space $\vec{r}$ is supposed to obey the natural exponential decay law:

$$S(\vec{r}, t) = e^{-kD(\vec{r},t)} \qquad (1)$$

where $k$ is Standard Survival Rate (m² /J), and $D(\vec{r}, t)$ is the irradiation dose (J/m²). Air turbulence and temperature are not considered since air turbulence effects are significantly less variable inside a Covid Clean irradiation chamber due to its lower volume and temperature needs further evaluation. These effects may be significant inside a space with a higher volume [5].

The average survival rate within the irradiation chamber of the volume *V* is:

$$S = \frac{1}{V} \iiint S(\vec{r}, t) d^3\vec{r} \qquad (2)$$

is usually applied in the form:

$$S(t) = e^{-kD(t)} \qquad (3)$$

According to [6], U.V. dose to be applied to reach 90% disinfection of a virus population (D90) is $D90 = \frac{\log(10)}{k}$. This parameter D90 allows to get the constant $k$ for any microorganism. The present calculations utilize the results obtained by Kowalski, Walsh, and Petraitis (2020) in their study [7] of the effects of UV light on coronaviruses, which are summarized in Table 1 below.

*Table 1. Summary of U.V. light studies on Coronaviruses [7].*

| Microbe | D90 dose (J/m²) | k (m²/J) |
|---|---|---|
| Coronavirus | 7 | 0.35120 |
| Berne virus (Coronaviridae) | 7 | 0.32100 |
| Murine Coronavirus (MHV) | 15 | 0.15351 |
| Canine Coronavirus (CCV) | 29 | 0.08079 |
| Murine Coronavirus (MHV) | 29 | 0.08079 |
| SARS Coronavirus CoV-P9 | 40 | 0.05750 |
| Murine Coronavirus (MHV) | 103 | 0.02240 |
| SARS Coronavirus (Hanoi) | 134 | 0.01720 |
| SARS Coronavirus (Urbani) | 241 | 0.00955 |
| *Average* | 67 | *0.03433* |

The spatial distribution of the survival fraction $S(\vec{r}, t)$ and UVGI dose may be found with the application of computational fluid dynamics (CFD). For example, an upper-room UVGI system in a naturally ventilated multi-bed hospital ward was simulated in [5]. The results obtained for a steady-state scenario in such a complicated 3D geometry were eventually converted into terms of average dose and cleaning factor shown in Table 2. Here $D_W$ is the volume averaged UVGI dose over the whole ward and is used as a single parameter to characterize the effectiveness of the UVGI devices over the entire space. $D_{BED}$ is the average UVGI dose calculated in a 0.75 × 0.70 × 2.4 m air volume directly above each bed and was selected as a parameter estimating the patient breathing zone. Below, the same notations are used in the calculations presented here.

*Table 2. Comparison of CFD predictions with mixing ventilation model [8].*

| Ventilation rate (ACH) | $D_w$ Mixing model (J/m²) | $D_w$ CFD Average (J/m²) | Air borne reduction due to U.V. (%) | Reduction due to ventilation w.r.t 2 ACH (%) | Total reduction w.r.t 2 ACH (%) |
|---|---|---|---|---|---|
| 2 | 4.629 | 4.06 | 64.9 | 0 | 64.9 |
| 6 | 1.543 | 1.35 | 38.2 | 66.7 | 79.4 |
| 12 | 0.771 | 0.727 | 23.6 | 83.3 | 87.3 |

In this study, the cylindrical symmetry of the model allows accounting of the distance-dependent radiation intensity. Putting Eq (1) into Eq (2) gives

$$S(t) = \frac{1}{V} \iiint S(\vec{r}, t) d^3\vec{r} = \frac{2}{r_2^2 - r_1^2} \int_{r_1}^{r_2} e^{-kI(r)t} \, r dr \tag{4}$$

where $r_1$ and $r_2$ are the UVGI bulb and irradiation chamber radius and length $L_{bulb}$ (m); time $t$ (s) of airflow within the irradiation chamber for one cycle. One cycle refers to one pass of air through a single UVGI-AP and is equal to $t_1$, and the correspondent value of one cycle survival rate is denoted as $S_1 = S(t_1)$ where

$$t_1 = \frac{V_{bulb}}{F_{rate}}, \tag{5}$$

with the bulb length $L_{bulb}$ and volume $V_{bulb} = \pi r_2^2 L_{bulb}$, the fan rate $F_{rate}$ (m³/s). Next, $I(r)$ is the intensity of ultraviolet C (UVC) radiation (W/m²) at the distance $r$ from the irradiation chamber axis. According to Gauss law, in the case $r_1 \ll L_{bulb}$ the formula approximately gives it:

$$I(r) = \varepsilon \rho \frac{P_{bulb}}{2\pi r \, L_{bulb}} \tag{6}$$

here $\varepsilon$ is the efficiency of the UVGI source or fraction of its electric power $P_{bulb}$ that goes into UVC radiation and is around 30%; $\rho$ is a coefficient that depends on the irradiation chamber inner wall reflectivity. The last calculation considers the material reflectance of the inner irradiation $R$. The value of $\rho$ increases by the multiple reflections inside the irradiation chamber. Its upper limit may be estimated as an infinite geometric series with the standard ratio $R$, $\rho = \frac{1}{1-R}$. In different cases of $R$, the increase of UVC radiation $\varepsilon \rho$ may be very large (see Table 3).

Any exponentially decaying process that has its correspondent decay time $\tau$ which typically is measured in experiments as the reverse slope of the logarithm of the survival fraction $\tau = \frac{t_1}{\ln S_1}$. Another interesting parameter of the process is the number of cycles needed to deactivate airborne biohazards up to 99.99% – a reduction to nearly zero virus concentration in an empty room.

*Table 3. U.V. reflectivity of standard materials.*

| Material | Reflectivity R | Boosting UV Power $\varepsilon\rho$ |
|---|---|---|
| Porous PTFE, [9] | 97% | 10 |
| e-PTFE, [10] | 95% | 6 |
| Aluminum – sputtered on glass [10] | 80% | 1.5 |
| Aluminum foil [10] | 73% | 1.1 |
| Stainless steel (various formulas) [10] | 20 – 28% | 0.38 – 0.42 |

Consider an equation of the number of viruses in the room time dependence $N(t)$. The rate of this value is the difference between relative incoming flow after irradiation $\frac{F_{rate}}{V_{room}} S_1 N(t)$ and outcoming into the chamber part $\frac{F_{rate}}{V_{room}} N(t)$:

$$\frac{dN(t)}{dt} = -\frac{F_{rate}}{V_{room}} (1 - S_1) N(t) \tag{7}$$

The solution of Eq (7) has the form:

$$\frac{N(t)}{N(0)} = \exp\left\{-\frac{F_{rate}}{V_{room}} (1 - S_1) t\right\} \tag{8}$$

In the case of 99.99% cleaning $\frac{N(t_c)}{N(0)} = 10^{-4}$, the correspondent time $t_c$ (in hours) and number of air changes $n_c$ are given by the following equation:

$$t_c = n_c \frac{V_{room}}{60 \, F_{rate}}, \quad n_c = \frac{4 \ln 10}{(1 - S_1)}. \tag{9}$$

Here $V_{room}$ is the room volume.

These indices, the one cycle survival rate $S_1 = S(t_1)$, number of cycles $n_c$, and time of 99.99% deactivation $t_c$, are prominent features of a Covid Clean UVGI-AP. However, it is not sufficient to produce clean air in an isolation room with a patient exhaling airborne biohazard such as a virus.

In the case of the steady state, the process may be described by the material balance equation. Suppose the breathing rate of a patient is varying in the range ~ 20 – 80 L/min or ~ 1 – 3 CFM, put its rate to $B_{rate} = 2$ CFM. The polluted airflow is equal to the product $B_{rate} C_{BED}$, where $C_{BED}$ is a concentration of viruses nearest the polluting source. In the steady state this flow is equal to the cleaning one. The cleaning flow is the velocity of $C_w$ decrease, where $C_w$ is the mean value of the virus concentration in the room.

$$B_{rate} \, C_{BED} = F_{rate} \, C_w \, (1 - S_1) \tag{10}$$

UVC deactivates viruses in the air of the irradiation chamber during the time $t_1$ that depends on the irradiation chamber volume and fan rate. The cleaning factor of the permanent process is

$$\frac{C_{BED}}{C_w} = \frac{F_{rate}}{B_{rate}}(1 - S_1) \qquad (11)$$

Thus, all the resulting parameters of the calculator are explained, except the apparent coefficients of unit conversions.

## Results and Discussions

Here, we present the typical results that the model produces. The effectiveness of electric to UVC radiation conversion varies in a wide range for different UVGI sources [11]. Here it was set to 30%, which is close to the minimal value. The bulb radius is set to $r_1 = 0.5"$. However, the value is not significant since the bulb occupies only a small fraction of the total volume inside the irradiation chamber. Figure 2 shows how the U.V. radiation intensity (in W/m²) decreases along the radial direction in the case of zero reflectivity $R = 0$ and $r_1 \ll L_{bulb}$. The curve starts from the border of the bulb surface $r_1$ and ends at the border of the irradiation chamber inner wall $r_2$ (dashed red lines). In the case of mirror walls, U.V. dispersion can be more uniform [10].

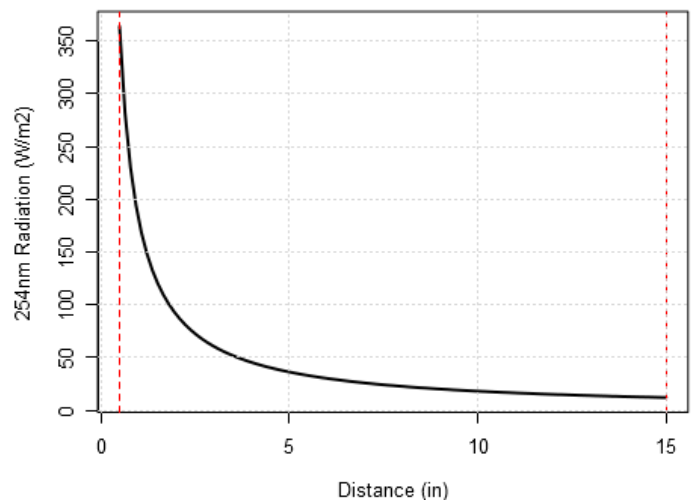

*Figure 2. Radiation power by distance from UVGI source in the case of zero reflectivity R = 0. The left dashed line indicates the border of the bulb surface r₁, and the right dashed line indicates the border of the irradiation inner wall r₂.*

The user interface of the calculator can vary all the other parameters of the model. Table 4 contains the parameters used to calculate the results of this section.

*Table 4. Parameters of the calculations which were chosen in the user interface of the calculator.*

| Parameter | Notation | Value | Unit |
|---|---|---|---|
| Target Virus D90 Dose (J/m2) | $D90$ | 241 | J/m² |
| 254nm UVGI Total Bulb Power | $P_{bulb}$ | 145 | W |
| Lamp Length | $L_{bulb}$ | 61 | inch |
| Irradiation Chamber Radius | $r_2$ | 15 | inch |
| Wall Reflectance | $R$ | 97%, 0 | |
| Air Flow | $F_{rate}$ | 135 | CFM |
| Room Volume | $V_{room}$ | 675 | CF |

In Table 5, the results of the most reflective wall are compared to ones calculated at zero reflectance. *One Cycle Survival Fraction of Selected Virus* is $S_1 = S(t_1)$ calculated at the time $t_1$ given by Eqs (4, 5). The last is *One Cycle Irradiation Time in Chamber*, the period of any virus under U.V. radiation inside the irradiation chamber. *Air Changes to Clean Empty Room To 99.99%* and *Time To Clean Empty Room To 99.99%* are the above mentioned $n_c$ and $t_c$, correspondingly, Eq (9). *Permanent Cleaning Factor Of Room With A Patient* refers to the ratio $\frac{C_{BED}}{C_W}$ given by Eq (11). And last, *Virus Decay Time $\tau$* and *Air Changes Per* Hour (ACH) are well-known parameters [12].

*Table 5. Result of the calculations in the cases of 97% and zero wall reflectivity.*

| Parameter | Value at R = 0 | Value at R = 97% | Unit |
|---|---|---|---|
| One Cycle Survival Fraction of Selected Virus | 15.925 | 0.000 | % |
| One Cycle Irradiation Time in Chamber | 10.715 | 10.715 | second |
| Air Changes to Clean Empty Room To 99.99% | 10.955 | 9.21 | |
| Time to Clean Empty Room To 99.99% | 0.913 | 0.768 | hour |
| Permanent Cleaning Factor of Room With A Patient | 56.75 | 67.5 | |
| Virus Decay Time | 5.832 | 0.241 | second |
| Air Changes Per Hour (ACH) | 12 | 12 | 1/hour |

The most reflective walls produce drastic changes in biohazard deactivation rates. The inactivation is many times more effective. The number of cycles to clean an empty room decreases by 25 times.

Figure 2 shows the survival curves of SARS Coronavirus (Urbani) at different distances from the UVGI source with and without reflective covering by Porous PTFE, [6]. It may be seen that 97% reflectance produces a considerable increase in UVGI performance. That reflectance amplifies 33 times the U.V. intensity, which is in the exponent in Eq (4). Note, however, that the calculation presumes strictly backward reflection that may not take place in a real device.

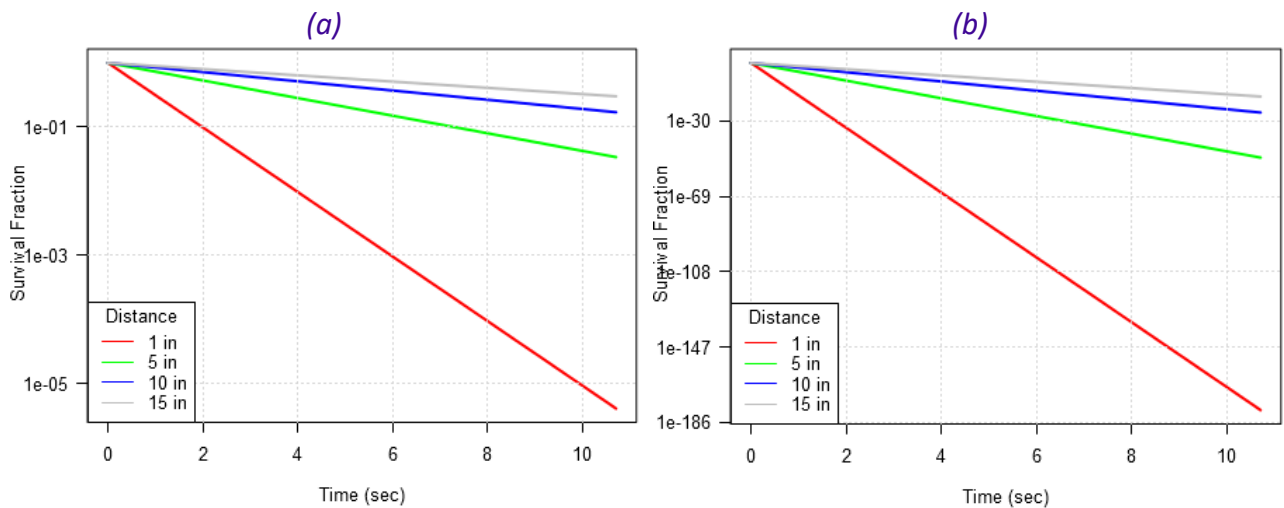

Figure 2. Survival curves for a set of distances in the case of zero reflectance (a) and Porous PTFE 97% (b).

## Conclusion

This paper describes the physical model and algorithm behind how the UVGI scientific calculator yields several useful results based on inputted parameters. Presented examples of the calculations illustrate how different factors influence various degrees of virus inactivation. Using reflective walls makes UVGI much more effective, significantly reduces the room cleaning time, and increases the efficiency of the permanent air cleaning process.

To summarize, the results suggest that the UVGI scientific calculator provides a quick and easy estimation of airborne biohazard deactivation and tracks trends and dynamic relations of the parameter inputs of the process.

The algorithm is implemented in the web application *UVGI Scientific Calculator* that has been used to facilitate the engineering and design of air-purifying equipment for hospitals and individuals.

In conclusion, this calculator shows that it is possible to clean air from within an isolation room and release it into the outer environment without needing to be vented out a window or into a hospital heating, ventilation, and air conditioning (HVAC) system. When air is pushed into HVAC systems from multiple rooms, there sometimes exists over pressurization and the risk for leaks of infectious air in the system downstream. Venting out a window is not always possible due to location such as at a sports stadium. This system provides a solution that is highly effective at isolating patients and staff from one another. The mobile and modular nature of the ultraviolet germicidal air-purifying temporary negative pressure isolation (UVGI-AP-TNPI) units leads to considerable global value in containing infectious diseases and reducing healthcare-associated infection which costs the U.S. alone 90,000 deaths per year and between $28-45-Billion annually [13].